\documentclass[english,aps,prb,twocolumn,showpacs]{revtex4}
\usepackage{amsmath}
\usepackage{amssymb}
\usepackage[dvipdfmx]{graphicx}

\newcommand{\bvec}[1]{\mbox{\boldmath $#1$}}

\makeatletter
\usepackage[colorlinks=true,urlcolor=blue,citecolor=blue,linkcolor=blue,breaklinks=true]{hyperref}
\makeatother
\begin{document}

\title{FLEX+DMFT approach to the $d$-wave superconducting phase diagram of the two-dimensional Hubbard model}

\author{Motoharu Kitatani}
\affiliation{Department of Physics, University of Tokyo, Hongo, Tokyo 113-0033, Japan}

\author{Naoto Tsuji\footnote[1]{Present address: RIKEN Center for Emergent Matter Science (CEMS), Wako 351-0198, Japan.}}
\affiliation{Department of Physics, University of Tokyo, Hongo, Tokyo 113-0033, Japan}

\author{Hideo Aoki}
\affiliation{Department of Physics, University of Tokyo, Hongo, Tokyo 113-0033, Japan}

\date{\today}
\begin{abstract}
The dynamical mean-field theory (DMFT) combined with the 
fluctuation exchange (FLEX) method, 
namely FLEX+DMFT, is an approach for correlated electron systems to 
incorporate both local and nonlocal long-range correlations in a self-consistent manner. 
We formulate FLEX+DMFT in a systematic way starting from a Luttinger-Ward functional,
and apply it to study the $d$-wave superconductivity in the two-dimensional repulsive Hubbard model.
The critical temperature ($T_c$) curve obtained in 
the FLEX+DMFT exhibits a dome structure as a function of 
the filling, which has not been clearly observed in the FLEX approach alone.  
We trace back the origin of the dome to the local vertex 
correction from DMFT that renders a filling dependence in 
the FLEX self-energy.
We compare the results with those of GW+DMFT, where the $T_c$-dome structure
is qualitatively reproduced due to the same vertex correction effect,
but a crucial difference from FLEX+DMFT is that $T_c$ is always estimated below the N\'{e}el temperature in GW+DMFT. 
The single-particle spectral function obtained with FLEX+DMFT exhibits 
a double-peak structure as a precursor of the Hubbard bands at temperatures above $T_c$.
\end{abstract}


\maketitle

\section{Introduction}
Despite a long history of physics of the high-$T_c$ cuprate,\cite{LBCO,scalapino_review}
we are still some way from a full understanding of the 
superconductivity. There is a general consensus that the 
supercurrent flows on each Cu-O plane, which can be modeled by 
the repulsive Hubbard model on the square lattice.  
There are actually two essential factors here: 
the repulsive Hubbard interaction can give rise to 
a pairing interaction in the $d$-wave channel 
mediated by antiferromagnetic spin fluctuations,\cite{RPA} while 
the very same 
interaction also introduces Mott's metal-insulator transition\cite{Mott} that hinders the superconductivity around half-filling for strong enough interactions.
Capturing these two features simultaneously 
still remains a theoretically challenging task.  
As numerical methods for treating the strongly correlated electron systems, 
there are the exact diagonalization and 
quantum Monte Carlo (QMC) methods \cite{QMC}
that are exact within numerical errors, 
but the former can only deal with limited system sizes, while 
the latter suffers from the sign problem.
  
However, we do have theoretical 
methods that can deal with each of the $d$-wave 
pairing and Mott's transition separately: Namely, 
we have on one hand 
the fluctuation-exchange (FLEX) approximation,\cite{FLEX} 
one of the perturbative methods for many-body physics that can describe 
the spin-fluctuation mediated $d$-wave pairing. On the other hand, we have 
the dynamical mean-field theory
(DMFT),\cite{d-infinite,DMFT,DMFT_review} 
which can describe the Mott transition.  
To be more precise, the FLEX describes the 
momentum dependence of the effective pairing interaction 
mediated by the antiferromagnetic spin fluctuations, 
which is essential for the anisotropic 
pairing,\cite{FLEX} but the method, being perturbative, cannot 
describe the Mott transition in the regime close to the half-filling. 
The DMFT, although mean-field theoretic, describes Mott's insulator 
in terms of the (non-perturbative) 
correlation effect that is local (i.e., momentum-independent) 
but dynamical (i.e., incorporating temporal fluctuations), 
and becomes exact in the limit of infinite spatial dimensions
of a lattice model.\cite{d-infinite}

There are many extensions of DMFT to
include momentum dependence of the
self-energy.\cite{DCA,cDMFT,DCA-Sakai,DCA-Gull,GW+DMFT2002,GW+DMFT2003,DgammaA,DgammaA-lambda,DF,DF2008,DF2009,DF2009_2,one-particle-irr,DMFFRG,MBPT+DMFT}
One is the cluster extension of DMFT,\cite{DCA,cDMFT}
which is employed, e.g., for explaining 
the pseudogap in the cuprates as a 
momentum-selective Mott transition.\cite{DCA-Sakai,DCA-Gull}
However, in practice it is quite hard in this scheme to attain large cluster sizes and 
to incorporate spatially long-ranged components in the self-energy
in a strongly correlated regime.
It is also computationally very demanding to treat the $d$-wave superconducting phase, or to extend to 
more complicated models such as 
multi-orbital systems with a large cluster size retained 
in the cluster DMFT.

More realistically, we have alternative and numerically feasible extensions of DMFT that combines DMFT with a certain resummation technique
of nonlocal self-energy diagrams,
such as
GW+(E)DMFT,\cite{GW+DMFT2002,GW+DMFT2003} D$\Gamma$A,\cite{DgammaA,DgammaA-lambda} and the dual-fermion approach.\cite{DF,DF2008,DF2009}  
These schemes can treat momentum-dependent self-energies describing nonlocal long-range correlations
with some selected diagrams taken into account. This has motivated us to 
take the FLEX+DMFT method,\cite{MBPT+DMFT,DMFT(FLEX)} where nonlocal FLEX diagrams are considered on top of DMFT local diagrams for the self-energy. 
We have opted for a method that evokes FLEX 
among other diagrammatic methods, since we are interested in the $d$-wave 
superconductivity 
mediated by antiferromagnetic fluctuations, 
which can be explicitly treated with FLEX.

In the present paper, we extend the FLEX+DMFT method to deal with 
the $d$-wave superconductivity in the two-dimensional repulsive Hubbard model, 
while a FLEX+DMFT has been applied to the normal phases of the Hubbard model in Ref.~\onlinecite{MBPT+DMFT}. 
To this end, we construct the Luttinger-Ward functional for FLEX+DMFT,
where double counting of local diagrams from FLEX and DMFT parts is unambiguously subtracted. 
Starting from the Luttinger-Ward functional formalism 
guarantees the conserved nature of DMFT (as well as FLEX) retained,
which is not always the case with other diagrammatic 
extensions of DMFT.
We then apply this FLEX+DMFT to the $d$-wave superconductivity in the two-dimensional repulsive Hubbard model
to obtain the superconducting phase diagram.

We find that the FLEX+DMFT result exhibits a 
$T_c$-dome structure of the superconducting phase diagram against band filling, 
which has not been observed in FLEX alone.  
We identify the origin of the dome to the local vertex 
correction from DMFT that renders a filling dependence in 
the FLEX self-energy.  
To elaborate this point, we compare this with the 
GW+DMFT method, in which 
only bubble diagrams are used to extend DMFT 
in considering a nonlocal self-energy correction, 
whereas both bubbles and ladders are included in FLEX+DMFT. The GW+DMFT result also exhibits a $T_c$-dome 
structure, 
but, unlike the FLEX+DMFT result, $T_c$ in  GW+DMFT is always below the N\'{e}el temperature, i.e., the antiferromagnetic order dominates over
$d$-wave superconductivity for the whole filling range. 
We have also obtained the single-particle spectral function with the FLEX+DMFT, which exhibits 
a double-peak structure above $T_c$ with a precursor of the Hubbard bands.

While the present scheme does not consider vertex corrections to the nonlocal ladder diagrams unlike the dual-fermion approach 
which is recently applied\cite{dual-fermion_superconductivity} to 
 the superconductivity in the Hubbard model, an advantage of the present 
method is that we define the Luttinger-Ward framework, which enables 
us to treat the normal self-energy and the anomalous ($d$-wave) 
self-energy on an 
equal footing as derivatives of the same Luttinger-Ward functional.

\section{FLEX+DMFT functional}
Let us formulate the FLEX+DMFT method by introducing a 
Luttinger-Ward functional $\Phi$,\cite{Luttinger-Ward} 
which basically consists of FLEX and DMFT diagrams. However, there is a double counting of local
self-energy diagrams between the two contributions, 
which must be subtracted.
We show that the double counting term is uniquely identified 
if one demands the conserving nature of the formalism.
Namely, we regard each of DMFT and FLEX as an approximation for 
the exact Luttinger-Ward functional of the dressed Green's function $G$ 
to propose a new functional,
in a manner similar to the GW+(E)DMFT scheme.\cite{GW+DMFT2002,GW+DMFT2003}
In DMFT, the approximate functional, $\Phi_{\rm DMFT}$, is the sum of
all types of the ring diagrams that only contain the local Green's function $ G_{\rm loc} $.  
On the other hand, the approximate functional in FLEX, $\Phi_{\rm FLEX}$, is the sum of specific (bubble and ladder) diagrams as shown in Fig.~1(a), 
which basically correspond to spin and charge fluctuations.

Then we can propose a functional in the FLEX+ DMFT scheme as 
\begin{equation}
	\Phi_{\rm FLEX+DMFT}[G] = \Phi_{\rm{DMFT}}[G_{\rm loc}]+\Phi_{\rm FLEX}[G]-\Phi_{\rm FLEX}^{\rm local}[G_{\rm loc}],
\end{equation}
where we have subtracted the local part of the 
FLEX functional $\Phi_{\rm FLEX}^{\rm local}[G_{\rm loc}]$
with $G_{\rm loc} = (1/N_{k}) \sum_{{\boldsymbol{k}}} G({\boldsymbol{k}})$ ($N_{k}$: number of $k$ points)
to avoid the double counting.
Since both $\Phi_{\rm DMFT}$ and $\Phi_{\rm FLEX}$ are expressed as functionals of 
dressed Green's functions, the overlap between the two is uniquely determined as a set of  
diagrams in the $\Phi_{\rm FLEX}[G]$ 
that {\it only} contain local dressed Green's functions,
which is nothing but $\Phi_{\rm FLEX}^{\rm local}[G_{\rm loc}]$.
We then obtain the self-energy in this scheme as a functional derivative,
\begin{align}
&
 \Sigma_{\rm FLEX+DMFT}[G] = \frac{\delta \Phi_{\rm FLEX+DMFT}}{\delta G} \notag \\
&=  \frac{\delta \Phi_{\rm{DMFT}}[G_{\rm loc}]}{\delta G_{\rm loc}} + \frac{\delta (\Phi_{\rm FLEX}-\Phi_{\rm FLEX}^{\rm local})}{\delta G}.
\label{selfenergy}
\end{align}
This way we retain the conserving nature of the 
approximation.\cite{Baym-Kadanoff,Baym} 
The first term in the last line of Eq.~(\ref{selfenergy}) is the local DMFT self-energy
$\Sigma_{\rm imp} \equiv \delta \Phi_{\rm{DMFT}}[G_{\rm loc}]/\delta G_{\rm loc}$.
The second term in Eq.~(\ref{selfenergy}), 
\begin{align}
\Sigma_{\rm FLEX}^{\rm nonloc} 
&\equiv 
\frac{\delta (\Phi_{\rm FLEX}- \Phi_{\rm FLEX}^{\rm local})}{\delta G} \nonumber \\
&= \Sigma_{\rm FLEX}[G]-\Sigma^{\rm loc}_{\rm FLEX}[G_{\rm loc}],
\label{nonlocal-selfenergy}
\end{align}
is the difference between the FLEX self-energy constructed from the lattice Green's function $G$ and that from the local Green's function $G_{\rm loc}$.
Note that $\Sigma_{\rm FLEX}^{\rm nonloc}$
contains some contributions from local parts of the self-energy, i.e.,
$\Sigma_{{\rm FLEX},ii}^{\rm nonloc}\neq 0$ ($i$: label of lattice sites).
For example, $\Sigma_{{\rm FLEX},ii}^{\rm nonloc}$
contains a diagram displayed in the left-hand side of Fig.~1(b) with $i\neq j$,
while a diagram shown on the right-hand side of Fig.~1(b) does not belong to $\Sigma_{{\rm FLEX},ii}^{\rm nonloc}$.

The self-consistency loop, which has to be a double loop in the present 
combined scheme, is depicted in Fig.~1(c): 
To start with, we define the DMFT mapping of a lattice model to an impurity model
in such a way that Green's function
of the mapped impurity model, $G_{\rm imp}$, coincides with the local Green's function 
for the original lattice model, $G_{\rm loc}$.
The local self-energy $\Sigma_{\rm imp}$
is calculated in the DMFT part of the self-consistency loop.  
The nonlocal part of the self-energy $\Sigma_{\rm FLEX}^{\rm nonloc}$
is then calculated in the FLEX loop.
We combine both $\Sigma_{\rm imp}$ and $\Sigma_{\rm FLEX}^{\rm nonloc}$ to obtain the full self-energy, from which 
we construct new (full and local) Green's functions.
We update each of them ($\Sigma_{\rm imp}$, $\Sigma_{\rm FLEX}^{\rm nonloc}$) alternately by using the corresponding loops until the whole loops 
[Fig.~1(c)] converge.

\begin{figure}
\begin{centering}
\includegraphics[width=1\columnwidth]{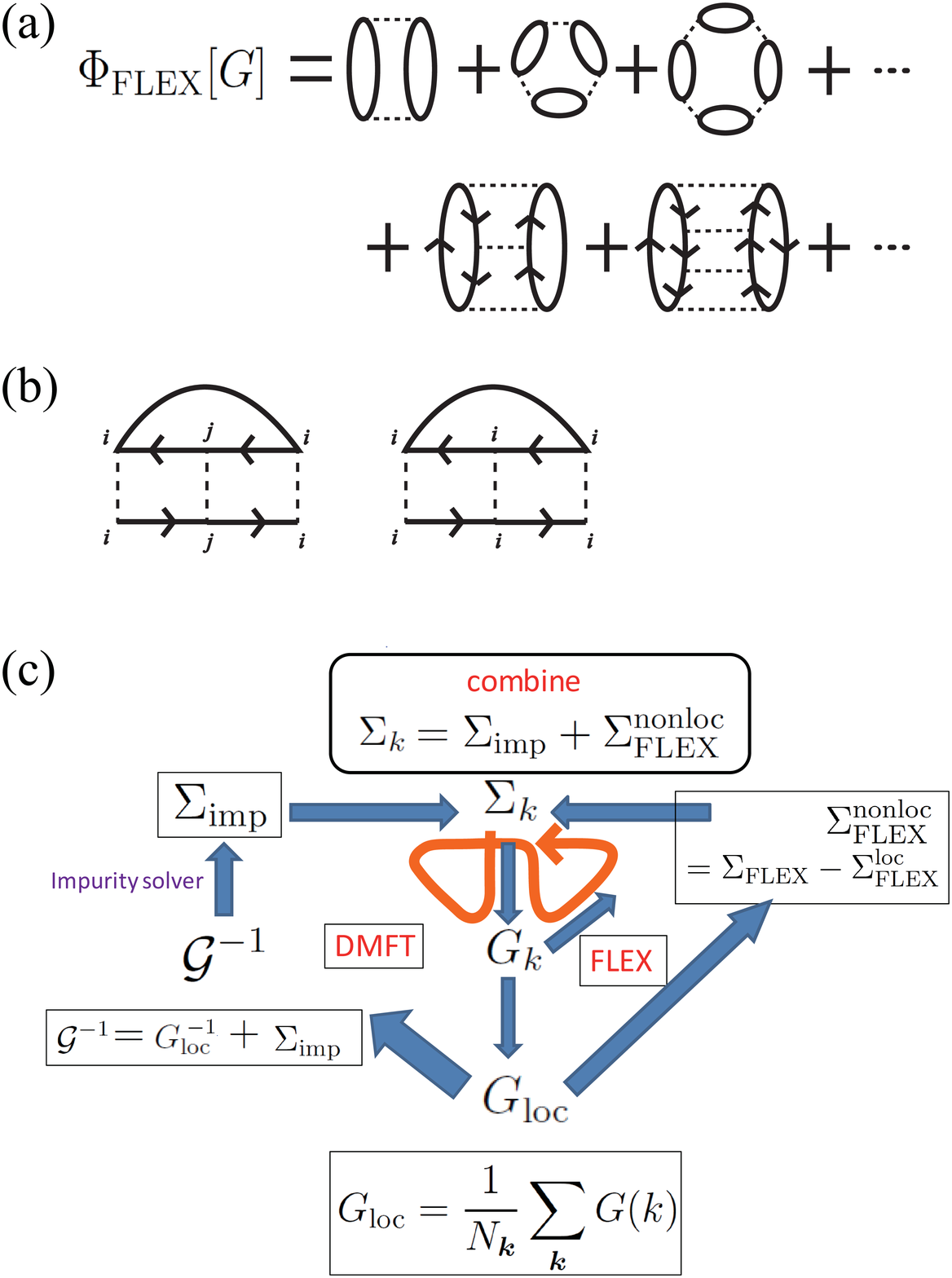}
\par\end{centering}
\caption{(Color online) (a) 
The Luttinger-Ward functional for FLEX. \cite{particle-particle}
(b) An example of a diagram contained in $\Sigma_{{\rm FLEX},ii}^{\rm nonloc}$ (left, with $i\neq j$) and omitted
 (right).
(c) Self-consistent double loops in the FLEX+DMFT formalism.
The algorithm flow for the loops is indicated by the red arrow.}
\end{figure}

The present scheme may be viewed as a new diagrammatic extension of the DMFT 
that incorporates vertex corrections  
into the (local part of) FLEX scheme.  
FLEX itself, being a perturbative method, is considered to become exact in the weak-coupling limit, 
while DMFT becomes exact in the atomic limit.  
Since FLEX+DMFT formalism here incorporates the functionals that dominate in 
either limit, 
it is expected to describe spin fluctuation effects and Mott's physics simultaneously.

\section{Application to the 2D Hubbard model}

Let us apply the FLEX+DMFT method to the repulsive Hubbard model on 
the square lattice, with a Hamiltonian, 
\begin{equation}
	{\cal H} = \sum_{{\boldsymbol{k}},\sigma} \epsilon({\boldsymbol{k}})c_{{\boldsymbol{k}},\sigma}^{\dag}c_{{\boldsymbol{k}},\sigma} + U \sum_{i} n_{i,\uparrow} n_{i,\downarrow}.
\end{equation}
Here $c_{{\boldsymbol{k}},\sigma}^{\dag}$ creates an electron in a 
Bloch state with wave-vector $ {\bvec k} = (k_{x},k_{y})$ and spin $ \sigma $, 
$U$ is the on-site repulsion, 
and $n_{i,\sigma} = c_{i,\sigma}^{\dag} c_{i,\sigma}$ is  the number operator. 
The two-dimensional band dispersion is given as
\begin{align}
	\epsilon({\boldsymbol k}) &= -2t({\rm cos} k_x + {\rm cos} k_y) \notag \\
							&\quad - 4t^{\prime} {\rm cos} k_x {\rm cos} k_y -2t^{\prime \prime}({\rm cos} 2k_x + {\rm cos} 2k_y)- \mu,
\end{align}
where $t,t^{\prime},$ and $t^{\prime \prime}$ represent the nearest-neighbor, second-neighbor, and
third-neighbor hoppings, respectively, while $ \mu $ is the chemical potential.
We shall compare the case with the nearest-neighbor hopping only ($t^{\prime}/t=t^{\prime\prime}/t=0$) with the case of 
$ t^{\prime}/t=-0.20 $, $ t^{\prime\prime}/t=0.16 $, 
which are the values estimated for a typical hole-doped, 
single-layered cuprate, 
HgBa$_2$CuO$_{4+\delta}$ with $T_c \simeq 90$K, 
with first-principles methods.\cite{nishiguchi_prb,sakakibara}  
Hereafter we take $|t|$ as the unit of energy.

In the single-band Hubbard model, the FLEX self-energy is computed as
\begin{align}
&
	\Sigma_{\rm FLEX}(k) = 
	\frac{1}{N_{\bm{k}} \beta} \sum_{k^{\prime}} \Bigl[
	\frac{3}{2}U^2 \frac{\chi_{0}(k-k^{\prime})}{1-U \chi_{0}(k-k^{\prime})} \notag \\
&
	+ \frac{1}{2}U^2 \frac{\chi_{0}(k-k^{\prime})}{1+U\chi_{0}(k-k^{\prime})}
	- U^2\chi_{0}(k-k^{\prime}) \Bigr] G(k^{\prime}),
\label{FLEX-selfenergy}
\end{align}
where $ \beta $ is the inverse temperature,
$k=(\omega_{n},{\boldsymbol k})$ with $\omega_{n}$ the fermionic Matsubara frequency, $G(k)$ is the Green's function, and
\begin{equation}
	\chi_{0}(q) = -\frac{1}{N_{\bm{k}} \beta}\sum_k G(k+q)G(k)
\label{chi}
\end{equation}
is the irreducible susceptibility.  
We can calculate $\Sigma^{\rm loc}_{\rm FLEX} $ by replacing $G$ with $G_{\rm loc}$ in Eqs.~(\ref {FLEX-selfenergy}) and (\ref {chi}).

To obtain $\Sigma_{\rm imp}$ in the DMFT procedure, 
we have to solve the impurity problem in DMFT.
Among various impurity solvers, here 
we adopt the modified iterative perturbation theory (modified IPT), where the original IPT is
modified for systems without particle-hole symmetry.\cite{mIPT}  The method is not computationally demanding, 
which facilitates a scanning 
over a wide parameter region 
to obtain the phase diagram, 
and also enables us to approach a region with 
large antiferromagnetic fluctuations where 
FLEX convergence critically slows down.
We have confirmed for various values of parameters 
that the continuous-time quantum Monte 
Carlo impurity solver \cite{CTQMC,CTQMC-review} implemented with the ALPS library\cite{ALPS1,ALPS2}
gives similar values for the eigenvalue of Eliashberg's equation even away from the half-filling.

When Green's function is obtained, we plug it into 
the linearized Eliashberg equation, 
\begin{equation}
	\lambda \Delta(k) = -\frac{1}{N_{\bm{k}}\beta}\sum_{k^{\prime}} V_{\rm eff}(k-k^{\prime})G(k^{\prime})G(-k^{\prime})\Delta(k^{\prime}).
\label{eliashberg}
\end{equation}
Here $\Delta(k)$ is the anomalous self-energy, which is
the gap function up to the renormalization factor, and
\begin{equation}
	V_{\rm eff}(k) = U + \frac{3}{2}U^2 \frac{\chi_0(k)}{1-U\chi_0(k)} - \frac{1}{2}U^2 \frac{\chi_0(k)}{1+U\chi_0(k)}
\end{equation}
is the effective pairing interaction (Fig.~2), 
where $\lambda$ is the eigenvalue for Eliashberg's equation, with superconducting transition identified as 
the temperature at which $\lambda = 1$.

\begin{figure}
\begin{centering}
\includegraphics[width=0.9\columnwidth]{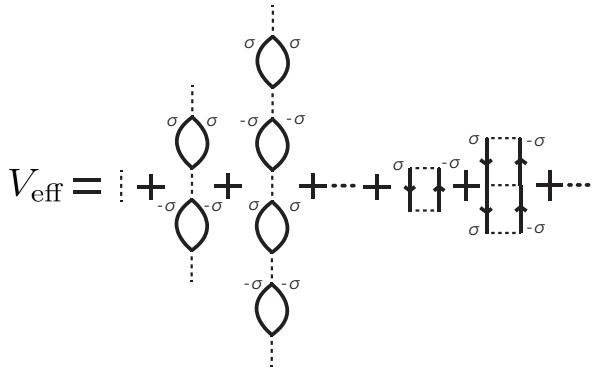}
\par\end{centering}
\caption{Effective pairing interaction in the FLEX.}
\end{figure}

At this point we should mention about 
the consistency of the approximate functional form $\Phi_{\rm FLEX+DMFT}$ and the linearized Eliashberg equation, Eq.~(\ref {eliashberg}).  
The Luttinger-Ward functional can be extended to incorporate the anomalous part, and the 
extended functional $\Phi[G,F^{\dag},F]$ is 
related to the anomalous self-energy through 
$\Delta = \delta \Phi/\delta F^{\dag}$, where $F^{\dag}$ is the 
anomalous Green's function, for which 
we should consider the local correction 
to the anomalous self-energy $\Delta$ 
as $\Delta_{\rm FLEX+DMFT} = \Delta_{\rm FLEX} + \Delta_{\rm loc}$ 
as  in Eq.~(\ref {selfenergy}) for the normal self-energy.  
Now, our interest here is the anisotropic, 
$d$-wave pairing instability in the repulsive model, 
for which we can ignore the local correction to the anomalous self-energy $\Delta_{\rm loc}$ which does not depend on momentum.
The remaining term $\Delta_{\rm FLEX}= \delta \Phi_{\rm FLEX}[G,F^{\dag},F]/\delta F^{\dag}$ is the
same as the right-hand side of the linearized Eliashberg equation (\ref {eliashberg}) if we linearize the anomalous part.\cite{FLEX_functional}  
Then our formalism treats the normal and anomalous self-energies consistently, 
as functional derivatives of the same Luttinger-Ward functional $\Phi_{\rm FLEX+DMFT}$.  
This is an advantage of using the Luttinger-Ward functional 
formalism in constructing a new scheme.  

\section{Results} \label{sec:Results}

\begin{figure}
\begin{centering}
\includegraphics[width=1\columnwidth]{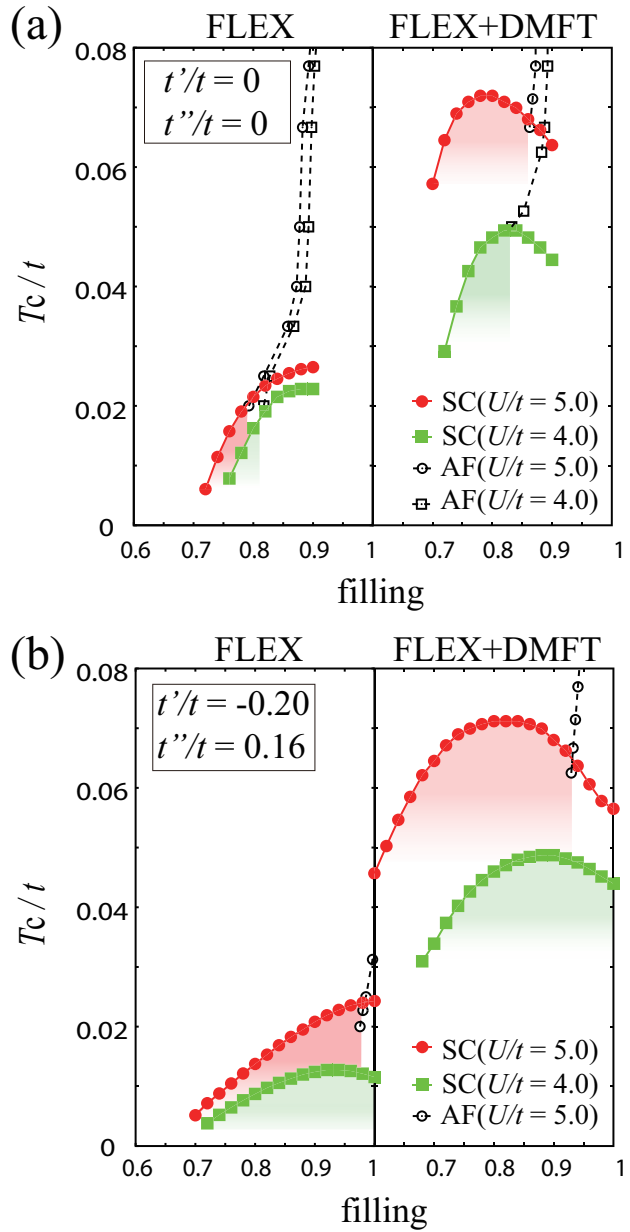}
\par\end{centering}
\caption{(Color online) Phase diagram against the band 
filling $n$ and the temperature $T/t$ in the FLEX+DMFT (right panels) as 
compared with FLEX (left). 
Here we take $U/t=4.0, 5.0$ and (a) $t^{\prime}/t=t^{\prime \prime}/t=0$, 
or (b) $t^{\prime}/t=-0.20, t^{\prime \prime}/t=0.16$. We also plot the AF phase boundaries (dotted lines) in the normal region, 
while color shading highlights the superconducting region with $T_c>T_{\rm AF}$ at each filling.
}
\end{figure}

We show the superconducting phase diagram of the two-dimensional Hubbard model obtained in the FLEX+DMFT in Fig.~3, right 
panels, where the FLEX result is also displayed in the left panels for comparison.  
We can immediately see that $T_c$ exhibits a dome structure in the FLEX+DMFT. This sharply contrasts with the FLEX result, 
where $T_c$ has been known to almost monotonically increase toward half-filling with some rounding off.\cite{review_yanase}
The presence of the $T_c$ dome in the FLEX+DMFT and its absence in the FLEX 
are seen for both the simple square lattice with $t^{\prime}=t^{\prime \prime}=0$ [Fig.~3(a)] and the case of $t^{\prime}=-0.20, t^{\prime \prime}=0.16$
[Fig.~3(b)].  

For the simple square lattice ($t^{\prime}=t^{\prime \prime}=0$), 
we cannot approach a region very close to half-filling 
because the antiferromagnetic (AF) fluctuations prevent 
the FLEX self-consistency loop from converging. 
For the same reason, it is difficult to attain convergence for systems with larger $U$.
As a measure of the AF order, 
we evaluated the AF phase boundaries (dashed lines in Fig.~3) 
determined from ${\rm max}_k[U\chi_{0}(k)] (=0.99$ here),\cite{AF} which is usually adopted in 
FLEX-type schemes to take account of the effect of the quasi-two-dimensional nature
(e.g., in three-dimensional layered systems),
although FLEX-type approaches are known to obey the Mermin-Wagner theorem 
that forbids finite-temperature AF phase transitions in an
isolated two-dimensional system.  
The estimated AF transition temperature $T_{\rm AF}$ becomes higher than
the superconducting $T_c$ as one approaches half-filling as shown in Fig.~3,
where the color shaded region indicates
the superconducting phase with $T_c>T_{\rm AF}$ (i.e., superconductivity dominating antiferromagnetism).  
The result suggests that a part of the $T_c$ dome 
is taken over by the AF phase
in the case of $t'=t''=0$ [Fig.~3(a), right] and $t'=-0.20, t''=0.16, U=5$ [Fig.~3(b), right]. For a smaller $U=4$, by contrast, 
we have an almost full $T_c$ dome with $T_c > T_{\rm AF}$ for 
$t^{\prime}=-0.20, t^{\prime \prime}=0.16$ [Fig.~3(b), right]. These are a key result in the present work.  

\begin{figure}
\begin{centering}
\includegraphics[width=1\columnwidth]{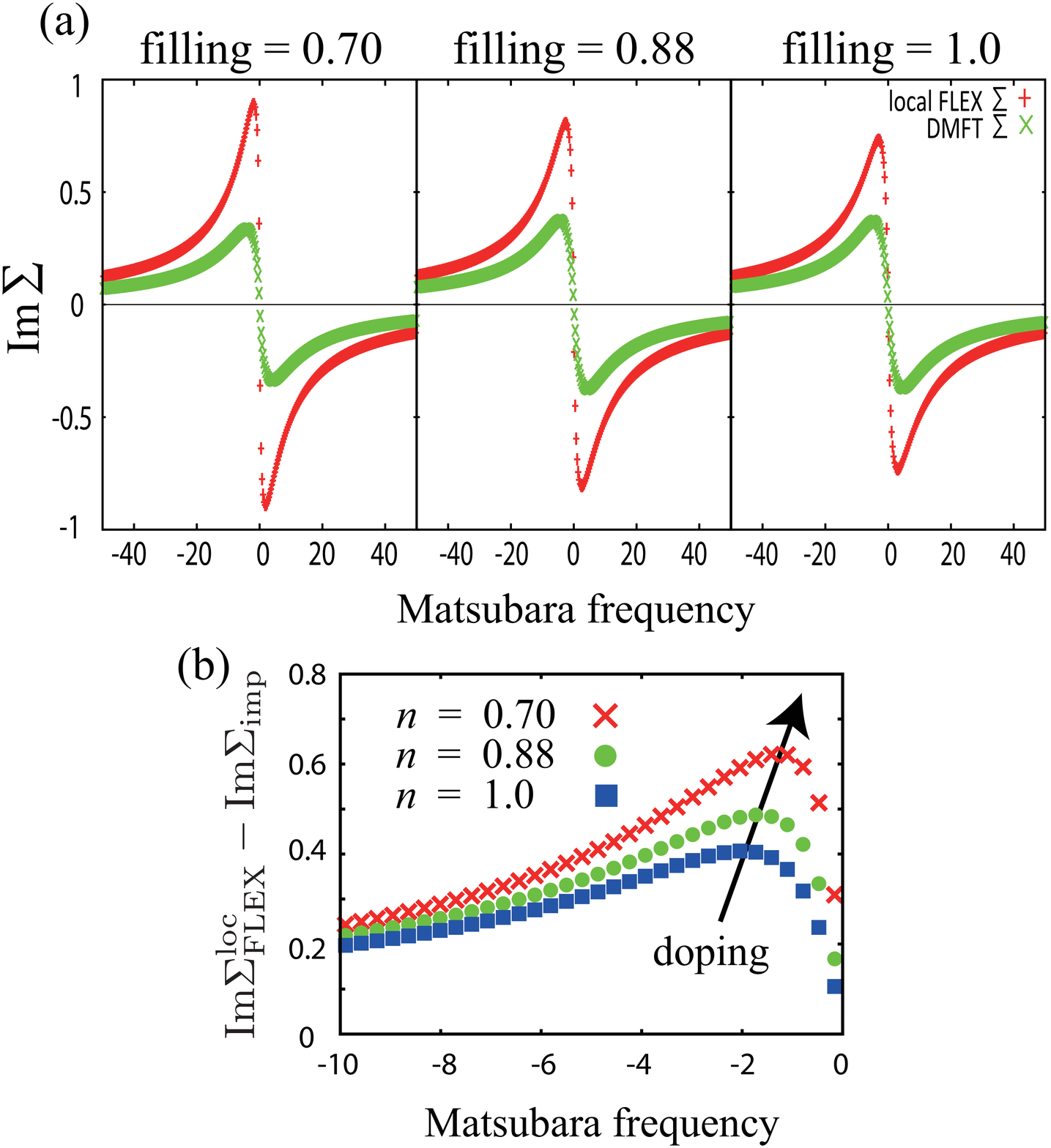}
\par\end{centering}

\caption{(Color online) (a) Comparison between the FLEX local self-energy ${\rm Im}\Sigma^{\rm loc}_{\rm FLEX}$ (red plus signs) 
and the DMFT self-energy ${\rm Im}\Sigma_{\rm imp}$ (green crosses) for the filling $n = 0.70$ (underdoped; left panel), $0.88$ (optimally doped; center), and $1.0$ (half-filled; right).  (b) 
The difference, Im$\Sigma^{\rm loc}_{\rm FLEX}-$Im$\Sigma_{\rm imp}$, for $n = 0.70$ (red crosses),
$0.88$ (green circles), and $1.0$ (blue squares).  
Here we take $U/t=4.0$, $\beta=20$, $t^{\prime}/t=-0.20$, and $t^{\prime \prime}/t=0.16$.}
\end{figure}

Now let us identify the physical 
origin of the appearance of the $T_c$ dome in the FLEX+DMFT.
In FLEX+DMFT, the self-energy is obtained from the FLEX and DMFT self-energies 
as $\Sigma_{\rm FLEX+DMFT}=\Sigma_{\rm FLEX}-(\Sigma_{\rm FLEX}^{\rm loc}-\Sigma_{\rm imp})$
[Eqs.~(\ref {selfenergy}) and (\ref {nonlocal-selfenergy})], i.e., a part of the local self-energy is replaced from that in FLEX with that in DMFT.  
Thus the quantity $\Sigma_{\rm FLEX}^{\rm loc}-\Sigma_{\rm imp}$ represents
the difference in the self-energy effect between FLEX and FLEX+DMFT.
We can actually take a look at 
$\Sigma^{\rm loc}_{\rm FLEX}$ and $\Sigma_{\rm imp}$, with fillings $n=0.7$ 
(underdoped), $0.88$ (optimally doped), and $1.0$ (half-filled) in Fig.~4
[the parameters are taken to be $U=4.0, \beta=20, t'=-0.20, t''=0.16$, which corresponds to
Fig.~3(b), right panel].
We first notice that the magnitude of the DMFT self-energy $\Sigma_{\rm imp}$
is smaller than that of FLEX $\Sigma_{\rm FLEX}^{\rm loc}$,
which means that the overestimation of the self-energy generally known to exist in FLEX is remedied in FLEX+DMFT
by the DMFT (local) vertex corrections.
More importantly, we can see that the difference, 
${\rm Im} \Sigma^{\rm loc}_{\rm FLEX}-{\rm Im} \Sigma_{\rm imp}$ [Fig.~4(b)], has a clear filling dependence, and it increases with doping.
Since $\Sigma_{\rm FLEX+DMFT}=\Sigma_{\rm FLEX}-(\Sigma_{\rm FLEX}^{\rm loc}-\Sigma_{\rm imp})$,
the result indicates that the reduction of the FLEX+DMFT self-energy due to DMFT correction is reduced as one approaches half-filling.
Thus  $T_c$ tends to be suppressed near half-filling as compared to that of FLEX
because of the {\it filling-dependent self-energy reduction} in FLEX+DMFT.
On the other hand, 
the pairing interaction itself arising from spin fluctuations 
becomes stronger toward half-filling due to better band nesting, as 
reflected in the FLEX result [Fig.~3, left panels] with $T_c$ almost monotonically increasing 
toward half-filling.  
Therefore, the FLEX+DMFT contains two factors with opposite 
filling dependencies, and 
we conclude that the $T_c$ dome in FLEX+DMFT arises 
from the {\it combined effect} of 
the nesting and filling-dependent self-energy reduction.

\begin{figure}
\begin{centering}
\includegraphics[width=1\columnwidth]{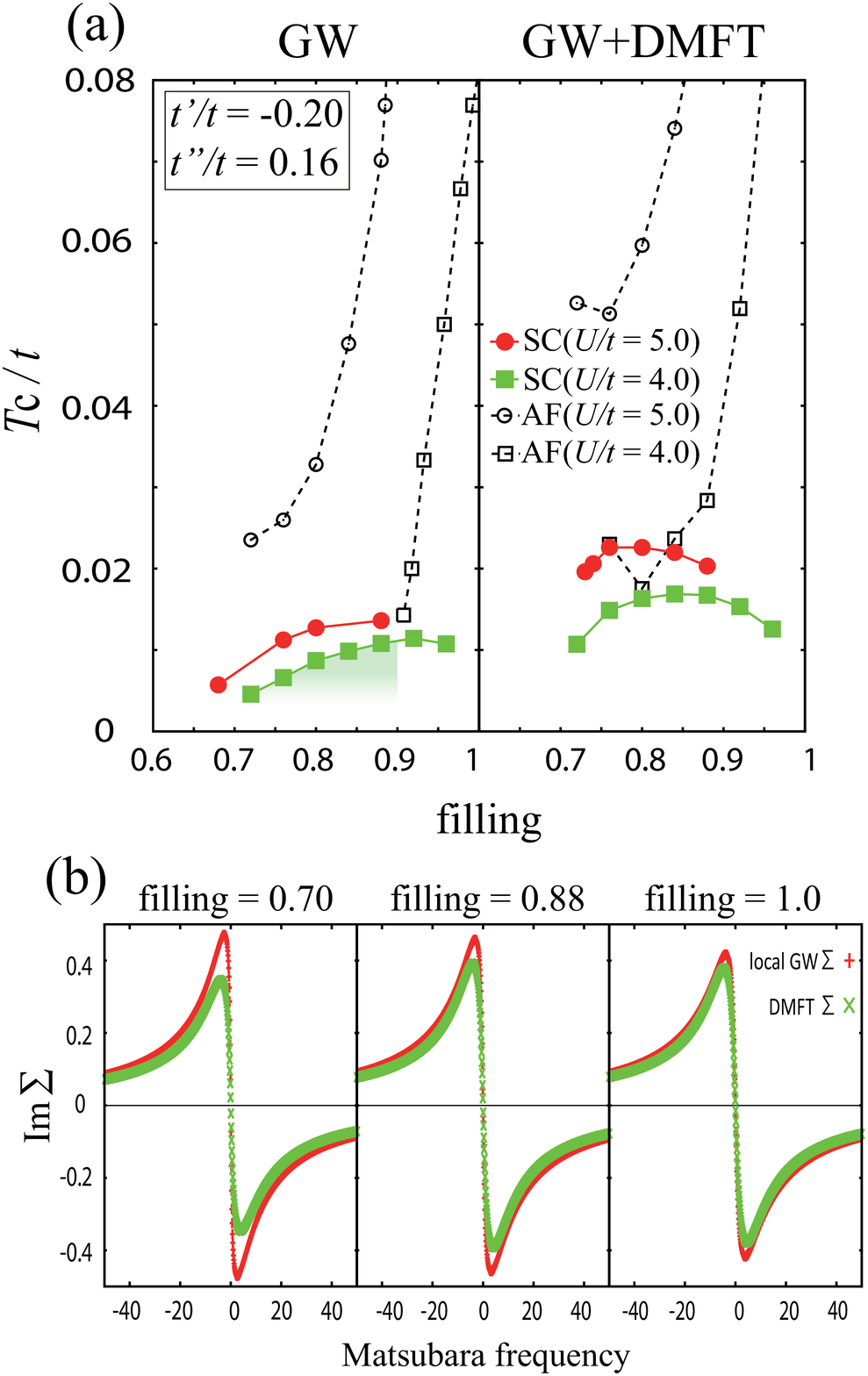}
\par\end{centering}
\caption{(Color online) (a) 
Phase diagram against temperature $T/t$ and band filling $n$ in GW+DMFT approximation (right) as compared with that in GW approximation (left) 
for $U/t=4.0$ (green squares) or $U/t=5.0$ (red circles).  
(b) Comparison between the GW local self-energy ${\rm Im}\Sigma^{\rm loc}_{\rm GW}$ (red plus signs)
and the DMFT self-energy ${\rm Im}\Sigma_{\rm imp}$ (green crosses) for the filling $n = 0.70$ (left panel), $0.88$ (center), and $1.0$ (right) with $U/t=4.0$ and $\beta=50$. 
Here we take $t^{\prime}/t= -0.20$, $t^{\prime \prime}/t=0.16$.  
We also plot antiferromagnetic (AF) phase boundaries (dotted lines) in the normal region, 
while color shading highlights the dome 
in the region where $T_c$ is above the AF boundary.
}
\end{figure}

In the FLEX+DMFT scheme, the self-energy reduction from DMFT takes place
only in the local part,
while the nonlocal self-energy is still considered to be overestimated, especially for ladder diagrams.\cite{MBPT+DMFT}
To examine this effect, we compare the present method with GW+DMFT,
where only the bubble diagrams are considered for 
the self-energy and the pairing interaction (whereas both bubbles and ladders are included in FLEX and FLEX+DMFT).
We show the GW+DMFT phase diagram, along with the GW result for comparison, in Fig.~5  for $t^{\prime}= -0.20$ and $t^{\prime \prime}=0.16$.  
We can see that, although 
the $T_c$ dome structure remains in the GW+DMFT result, 
$T_c$ is much reduced from the result of FLEX+DMFT.  
On the other hand, the AF transition temperature is much higher in  GW+DMFT than 
that of FLEX+DMFT. This makes the region in the dome 
where $T_c > T_{\rm AF}$  
[highlighted with color shadings in Fig.~5(a)] 
very narrow in  GW+DMFT.  In fact, 
for $t^{\prime}=t^{\prime \prime}=0$ the AF instability
becomes so strong that we cannot even obtain superconducting phase boundaries for the whole region of the fillings considered.

In Fig.~5(b), we display the GW local self-energy ${\rm Im}\Sigma^{\rm loc}_{\rm GW}$ 
as compared with the DMFT self-energy ${\rm Im}\Sigma_{\rm imp}$  for the filling $n = 0.70, 0.88, 1.0$ with $U=4.0$ and $\beta=50$.
We can see that the filling dependence is similar to those in the FLEX+DMFT in that the difference between the two self-energies increases with the doping. 
Hence we can conclude that the existence of the $T_c$ dome
is not an artifact in FLEX+DMFT, but is robust in both FLEX+DMFT and GW+DMFT
arising due to the same local vertex correction effect.
The overestimation of nonlocal self-energy thus does not affect the existence of the $T_c$ dome itself.  

The reason that the magnitude of $T_c$ is much smaller in GW+DMFT than 
in FLEX+DMFT is 
because ladder diagrams describing 
spin fluctuations are not taken into account in GW+DMFT.  
In this sense, GW+DMFT is closer to the mean-field theory than FLEX+DMFT,
which is also reflected in the higher AF transition temperature in GW+DMFT. 
Concomitantly, the pairing interaction mediated by spin fluctuations is reduced, which acts to reduce the superconducting $T_c$
in GW+DMFT rather than in FLEX+DMFT.
The fact that $T_c$ is always estimated below the N\'{e}el temperature in GW+DMFT suggests that
GW+DMFT underestimates the spin fluctuation effect, 
and is not enough to describe the $d$-wave superconductivity mediated by spin fluctuations.
Since the overestimated nonlocal self-energy in FLEX+DMFT is remedied in GW+DMFT,
the accurate estimation of the spin-fluctuation effect is expected to lie between
GW+DMFT and FLEX+DMFT.

To see the strength of local correlation, we measure the double occupancy, 
\begin{equation}
	\left< n_{\uparrow}n_{\downarrow} \right> = \frac{1}{U} {\rm Tr}(\Sigma G).
\end{equation}
We can see in Fig.~6 that the double occupancy 
becomes negative in the overdoped regime in FLEX, while this unphysical behavior is improved in FLEX+DMFT.  
We can regard this as another of the self-energy reduction effects: FLEX overestimates the correlation effect, while this is corrected 
by combining it with the DMFT.  A similar tendency is observed between GW and GW+DMFT, but the difference in the double occupancy is smaller.
This should be because the self-energy reduction effect is smaller [see Figs.~4(a) and~5(b)].

\begin{figure}[t]
\begin{centering}
\includegraphics[width=0.8\columnwidth]{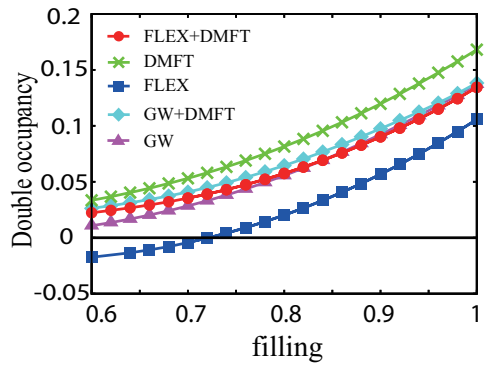}
\par\end{centering}
\caption{(Color online) Double occupancy against band filling $n$ in the FLEX+DMFT (red circles), DMFT (green crosses), FLEX (blue squares), GW+DMFT (light blue diamonds), and GW (purple triangles).
Here we take $U/t=4.0$, $\beta=20$, $t^{\prime}/t=-0.20$, and $t^{\prime \prime}/t=0.16$.}
\end{figure}

Let us finally examine the spectral function, 
which is calculated via analytical continuation with the Pad\'{e} approximation.  
The results for FLEX+DMFT, FLEX, GW, GW+DMFT, and DMFT 
at various fillings are shown in Fig.~7.
We can see that the filling dependence is similar among 
FLEX, GW, and DMFT in that we have a single peak that slightly shifts and broadens as we approach the half-filling.  
By contrast, in the method that combines DMFT with either FLEX or GW, the spectral function acquires a stronger filling dependence, where a 
marked double peak is observed at half-filling.  
Similar double-peak structures have been reported in the dual-fermion method\cite{DF2009_2} as an antiferromagnetic pseudogap, where the appearance of the 
double peak is consistent with the QMC result. 
Thus we can see that the interplay of the local and nonlocal long-range correlation effects
in FLEX+DMFT and GW+DMFT gives the 
double peak, which is considered to be a precursor of the Hubbard bands
with two peaks separated by about $U$, while the system is metallic.

\begin{figure}[t]
\begin{centering}
\includegraphics[width=1\columnwidth]{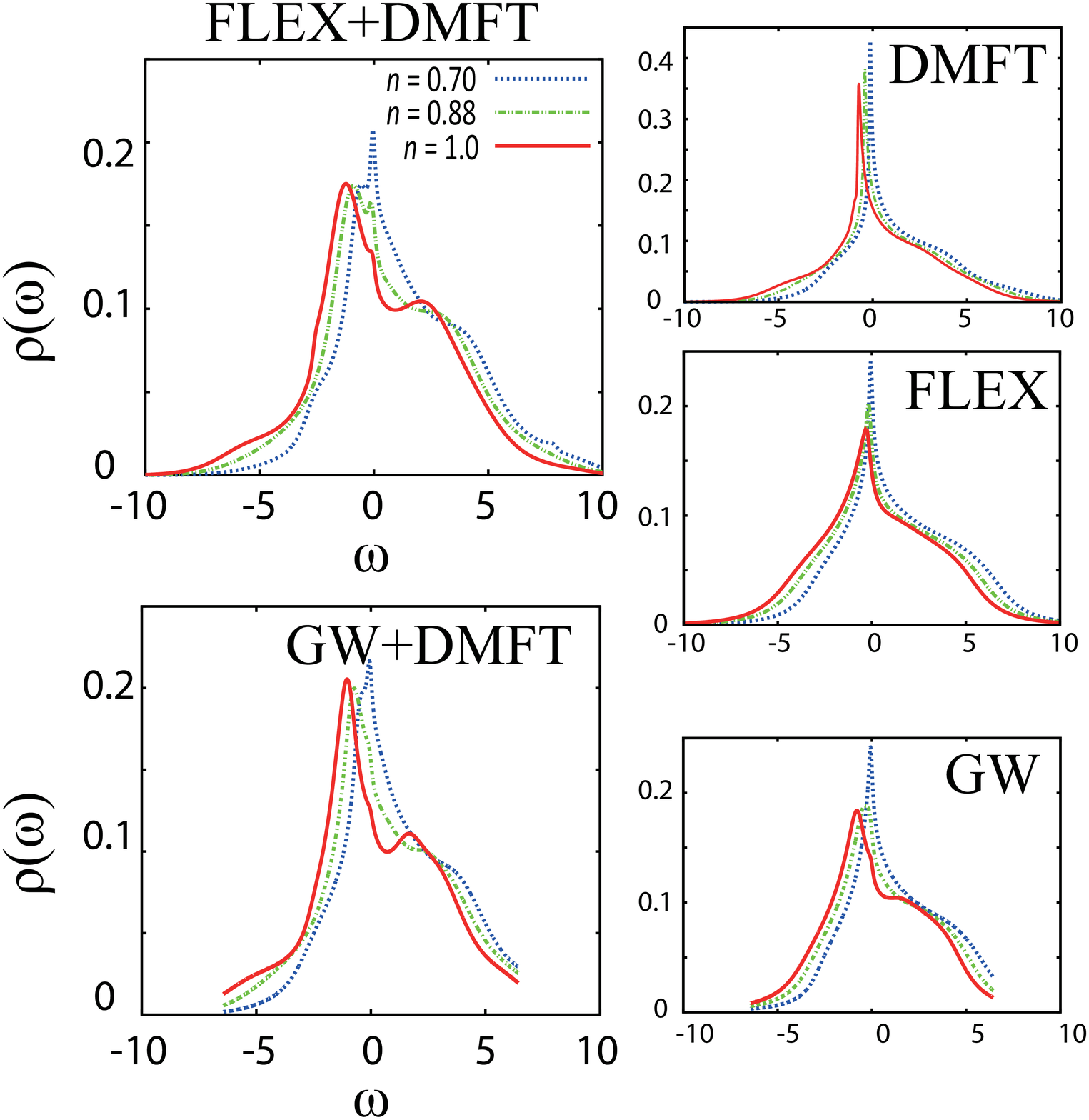}
\par\end{centering}
\caption{(Color online) Spectral functions at filling $n=0.70$ (underdoped; dotted lines), 0.88 (optimally doped; green), 1.0 (half-filling; red) in the FLEX+DMFT (top left) and GW+DMFT (bottom left) are 
compared with those in DMFT (top right), FLEX (middle right), and GW (bottom right). Here we take $U/t=4.0$, $\beta=20$, $t^{\prime}/t=-0.20$, and $t^{\prime \prime}/t=0.16$.}
\end{figure}

If we look more closely at the momentum-resolved spectral function $A({\boldsymbol k},\omega)$, we observe that there is a region in 
$k$ space near the Fermi energy where the 
spectral weight becomes slightly negative.
This might not be specific to the present method, since many extensions of 
DMFT do not guarantee positive-definite spectral weights.\cite{DF2009,one-particle-irr}
Since the magnitude of the negative part is negligibly small ($< 1\%$)  in the present case 
and this tends to occur in the overdoped regime, 
this does not affect the phase diagram and the density of states [Figs.~3 and 7] 
in the underdoped regime.  

\section{Summary and discussions}
We have employed the FLEX+DMFT approach in terms of a new Luttinger-Ward functional
to study the superconductivity in correlated electron systems.
This scheme is a diagrammatic extension of the DMFT, so that it can describe the $d$-wave superconductivity arising from $k$-dependent pairing interaction.  The scheme, being formulated in terms of the 
Luttinger-Ward functional, also has a virtue of the normal and anomalous self-energies treated on an equal footing.  
We have applied the FLEX+DMFT to the repulsive Hubbard model on the square lattice.  
We have found that FLEX+DMFT describes a $T_c$ dome structure, whose physical origin is traced 
back to a combination of opposite effects: 
The self-energy effect introduced in the FLEX+DMFT 
suppresses the superconductivity more strongly toward the half-filling 
due to the local-correlation effect, while spin fluctuations become stronger toward the half-filling due to band 
nesting.  
We also compare the FLEX+DMFT result with the GW+DMFT result, 
which reproduces the dome structure. This indicates 
that the dome is not an artifact of the overestimated nonlocal self-energy in 
FLEX+DMFT.  

Another observation is that there is a 
Pomeranchuk instability into electronic states with broken tetragonal symmetry
in the case of 
$t^{\prime}=-0.20,t^{\prime \prime}=0.16$ in both FLEX+DMFT and GW+DMFT.  
In this case, solutions with the four-fold rotationally symmetric Fermi surface become unstable, where we end up with 
a solution that breaks this symmetry 
when we start the calculation from an asymmetric initial input. 
While this instability is interesting in its own right, 
we have concentrated on the symmetric case in this study, 
and leave the analysis of the Pomeranchuk instability to another publication.

In order to improve the scheme to suppress the overestimation of the nonlocal FLEX self-energy,
we should consider the screening effect in the FLEX self-energy. 
For example, the two-particle self-consistent
method \cite{TPSC} takes account of vertex correction effects  
by considering the sum rule for the susceptibility, while a 
similar technique is also used to reduce the overestimated
spin fluctuations and their effect on the self-energy in D$\Gamma$A.\cite{DgammaA-lambda}
We expect these techniques bring some improvement to the present theory.

\section{Acknowledgement}
The authors wish to thank K. Nishiguchi for providing a FLEX coding and fruitful discussions, and J. Otsuki and H. Hafermann for useful discussions.
The present work was supported by a Grant-in-Aid for Scientific Research (Grant No. 26247057) from MEXT, M.K. was supported by the
advanced leading graduate course for photon science (ALPS), and N.T. was supported by a Grant-in-Aid for Scientific Research (Grant No. 25800192) from JSPS.

\newcommand{\PR}[3]{Phys. Rev. \textbf{#1},#2 (#3)}
\newcommand{\PRL}[3]{Phys. Rev. Lett. \textbf{#1},#2 (#3)}
\newcommand{\PRA}[3]{Phys. Rev. A \textbf{#1}, #2 (#3)}
\newcommand{\PRB}[3]{Phys. Rev. B \textbf{#1}, #2 (#3)}
\newcommand{\JPSJ}[3]{J. Phys. Soc. Jpn. \textbf{#1}, #2 (#3)}
\newcommand{\arxiv}[1]{arXiv:#1}
\newcommand{\RMP}[3]{Rev. Mod. Phys. \textbf{#1}, #2 (#3)}


\begin{thebibliography}{99}
\bibitem{particle-particle}
While we do not take account of the particle-particle ladder diagrams for the FLEX self-energy in this paper, 
we have checked that the result remains almost the same when we consider the diagrams, especially when we combine those with DMFT in the present FLEX+DMFT.
\bibitem{AF} The present estimate of the AF susceptibility is only on the mean-field level, while 
for a more precise treatment one should solve the Bethe-Salpeter equation by using $\delta ^2 \Phi/\delta G^2$ as the interaction kernel,
which will somewhat reduce $T_{\rm AF}$ for (quasi-)two dimensional systems.
\end{thebibliography}
\end{document}